%% file: AVGAN.tex
\documentclass[10pt,twocolumn,letterpaper]{article}

\usepackage[pagenumbers]{wacv} 

\usepackage{graphicx}
\usepackage{amsmath}
\usepackage{amssymb}
\usepackage{booktabs}

\usepackage{times}
\usepackage{epsfig}
\usepackage{subcaption}
\usepackage{multirow}
\usepackage{musicography} 
\usepackage{tikz}
\usepackage{xcolor} 
\usepackage{adjustbox} 
\usetikzlibrary{arrows, positioning, shapes, fit, calc, arrows.meta}

\usepackage{tablefootnote, footnotehyper} 

\usepackage[pagebackref,breaklinks,colorlinks]{hyperref}

\usepackage[capitalize]{cleveref}
\crefname{section}{Sec.}{Secs.}
\Crefname{section}{Section}{Sections}
\Crefname{table}{Table}{Tables}
\crefname{table}{Tab.}{Tabs.}

\def\wacvPaperID{1731} 
\def\confName{WACV}
\def\confYear{2024}

\begin{document}

\title{Listen and Move: Improving GANs Coherency in Agnostic Sound-to-Video Generation}

\author{Rafael Redondo\\
Eurecat, Centre Tecnològic de Catalunya, Tecnologies Multimèdia\\
Barcelona, 08005, Spain\\
{\tt\small rafael.redondo@eurecat.org}
}
\maketitle

\newcommand\rnncolor{magenta!60}
\newcommand\gencolor{blue!35}
\newcommand\discolor{rgb,255:red,252; green, 214; blue, 125}
\newcommand\lmscolor{rgb,255:red,75; green, 198; blue, 185}

\tikzstyle{stycat}=[circle, draw=black, fill=white, minimum size=0cm, font=\tiny, inner sep=0pt]
\tikzstyle{stysum}=[circle, draw=black, fill=white, minimum size=0cm, font=\scriptsize, inner sep=-1pt]

\newcommand{\icol}[1]{\left[\begin{smallmatrix}#1\end{smallmatrix}\right]}
\newcommand{\mbs}[1]{\boldsymbol{#1}}
\newcommand\mdoubleplus{\ensuremath{\mathbin{+\mkern-10mu+}}}
\newcommand\pms{\pm\scriptstyle}

\begin{abstract}
    Deep generative models have demonstrated the ability to create realistic audiovisual content, sometimes driven by domains of different nature. However, smooth temporal dynamics in video generation is a challenging problem. This work focuses on generic sound-to-video generation and proposes three main features to enhance both image quality and temporal coherency in generative adversarial models: a triple sound routing scheme, a multi-scale residual and dilated recurrent network for extended sound analysis, and a novel recurrent and directional convolutional layer for video prediction. Each of the proposed features improves, in both quality and coherency, the baseline neural architecture typically used in the SoTA, with the video prediction layer providing an extra temporal refinement.
\end{abstract}

\section{Introduction}
\label{sec:intro}

The invention of Generative Adversarial Networks (GANs) \cite{goodfellow2014generative} and later convolutional GANs \cite{radford2015unsupervised} has enabled the synthesis of a large and realistic variety of images. By confronting two networks, the generator and discriminator are able to implicitly learn complex data distributions. Shortly thereafter, conditional GANs devised new modulation mechanisms to have more control over the adversarial synthesis, by either conditional embeddings \cite{reed2016generative, zhang2017stackgan} or explicit conditional images \cite{isola2017image}.

The emergence of conditionals GANs resulted in a diversity of cross-modal synthesis
by mapping data between different domains. For that, a semantic link must exist between both source and target domain, \ie their probability distributions must not be independent. 
Some examples are text-to-image \cite{zhang2017stackgan, xu2018attngan, Zhu2019DMGANDM, Liang2020CPGANCG, Zhang2021CrossModalCL}, 
text-to-video \cite{pan2017create, tulyakov2018mocogan, balaji2019conditional, deng2019irc, kim2020tivgan, han2022show}, 
speech-to-face \cite{Vougioukas2019RealisticSF, Chen2019HierarchicalCT}, 
video-to-sound \cite{Li2018VideoGF, zhou2018visual, chen2020generating, chen2018visually}, 
or video-to-video\cite{wang2018video}.

This paper focuses on generic sound-to-video translation.
This entails a highly unconstrained problem, more than other modalities, because the physically plausible motion of objects has to be estimated exclusively according to the semantics of the produced sound.
Certainly, multiple visual variations are plausible for a single sound. Moreover, video synthesis must deal with pixel jittering and audio-visual lagging, which is irrelevant for still images and to which human perception is extremely sensitive.
In contrast to other modalities, such as speech-to-face, no prior distribution is assumed for either image or sound. Compared to pose-based approaches, this could tackle situations in which the human pose is difficult to infer, such as a face profile, occluding objects, or even an entire music ensemble performance. In a more general sense, it enables applications such as music-guided video restoration, audiovisual performance reinterpretation, agnostic speech replacement, or audio-reactive video animations.

\section{Related works}
\label{sec:related}

In recent years, the ability of GANs to translate from audio to image has been demonstrated \cite{chen2017deep, hao2018cmcgan, Wan2019TowardsAT, yang2020diverse, duan2021cascade}, yet frames are independently synthesized without temporal coherency.
A common practice to leverage GANs from image to video is to use some sort of 3D video discriminator \cite{saito2017temporal, Li2018VideoGF, Song2019TalkingFG, tulyakov2018mocogan, balaji2019conditional, deng2019irc, 
Zhou2019TalkingFG, kim2020tivgan, Song2019TalkingFG, Chen2019HierarchicalCT, Vougioukas2019RealisticSF}. Although this should ideally be sufficient, an image discriminator on its simpler layout improves the convergence in adversarial training, providing feedback for both image quality and temporal dynamics \cite{tulyakov2018mocogan}.

Another common approach to induce temporal coherency  
consists in feeding the generator with a series of noise vectors temporally encoded by a Recurrent Neural Network (RNN). Intuitively, the RNN maps a sequence of feedforward independent random variables to a sequence of correlated random variables, representing temporal dynamics along the video stream.

In particular, MoCoGAN \cite{tulyakov2018mocogan} ---recently extended in \cite{skorokhodov2022stylegan}--- proposed a novel framework for unconditional video generation disentangling motion from content by feeding the generator with an additional source of noise. The authors showed a decomposed representation was able to fabricate more coherent videos. Although it was not particularly adopted by other subsequent works, it is adapted in this work for audio conditioning. 

Alternatively, MoCoGAN-HD \cite{tian2021a} presented a particular feedforward scheme to map motion trajectories in a latent space pre-trained for still images, which ultimately fails to represent fine-grained temporal dynamics. Similarly, a text-driven sound-to-video method has been proposed in \cite{Lee2022SoundGuidedSV}.

Furthermore, Dilated RNNs have been used for raw audio generation \cite{Oord2016WaveNetAG} and data series processing \cite{chang2017dilated, Zhao2019RecurrentNN}. However, dilated schemes has been omitted for GANs video generation hitherto.


In parallel, some optimizations have been proposed in speech-driven video synthesis, specially tailored for facial attributes, involving mouth generation \cite{Chen2018LipMG, Song2019TalkingFG}, lip reading \cite{Song2019TalkingFG}, encoded action units \cite{Pumarola2018GANimationAF, Pham2017SpeechDriven3F}, or facial landmarks \cite{Chen2019HierarchicalCT}, a word-based learning for lip-facial attribute disentangling \cite{Zhou2019TalkingFG}, or 
a third discriminator for audio-visual synchronization \cite{Vougioukas2019RealisticSF}.
Most of these approaches use a reference frame, which reduces image uncertainty but ultimately produces rigid, unnatural movements.
Other approaches have exploited body landmarks to assist in music-driven video performances \cite{zhu2021let}. 
However, albeit significant, these methods cannot be applied in its entirety for generic (faceless or bodyless) sound-to-video synthesis.

To induce temporal coherency at video level, vid2vid \cite{wang2018video} relies on pre-trained flow models and previously synthesized frame feedback. Such an approach however is simply not feasible for sound-to-video.
Fortunately, the apparition of ConvRNNs \cite{Shi2015ConvolutionalLN, stollenga2015parallel}
has enabled an efficient spatial analysis and temporal coherence by replacing with convolutions the gated matrix multiplications inside the recurrent cell.
Vanilla ConvRNNs have proven to be more efficient than flow-based solutions for video prediction in non-causal configurations \cite{Byeon2018ContextVPFC}.
They have also been used for video encoding \cite{Ballas2016DelvingDI} and video representations 
\cite{Liang2017DualMG}, 
as well as landmark-based video synthesis from speech \cite{Chen2019HierarchicalCT} or music \cite{zhu2021let}. Inspired by \cite{Byeon2018ContextVPFC} to enhance temporal coherency, a novel directional and causal ConvGRU for video prediction is proposed in this work.

In addition, increasing image quality and resolution is particularly challenging in GANs. The stability of the adversarial training is usually affected as resolution scales. While most domain translation approaches use a two-stage architecture \cite{zhang2017stackgan, Wang2018HighResolutionIS, yang2020diverse, duan2021cascade}, an improved StyleGAN \cite{karras2020analyzing} has evidenced the benefit of either residual or skip connections.
Other methods have proposed different convolutional optimizations \cite{Kahembwe2020LowerDK, saito2020train}.
Still, the task of generating smooth high-quality videos is a challenging problem \cite{Acharya2018TowardsHR, saito2017temporal, tulyakov2018mocogan, Yushchenko2019MarkovDP, saito2020train, tian2021a}. 
The caveat, as shown later, is that sound-driven learning has its own particularities to scale up.

Finally, it is worth mentioning methods based on attentional architectures (VQGAN) for sound-to-video \cite{ge2022long, chatterjee2020sound2sight, le2021ccvs}, which however usually operate at low resolution and reveal collapsed examples.
Beyond GANs, AD-NeRFs \cite{guo2021ad} showed outstanding speech-to-video synthesis, but based on a facial-tailored solution separately modeling various bust parts.


\medskip\noindent\textbf{Contributions}.
The aim of this work is to improve both image quality and temporal coherency of generic sound-to-video GANs through three main novelties:

\begin{itemize}
    \item A more versatile triple sound routing for motion encoding, content representation, and conditional normalization layers.
    \item A residual multi-scale DilatedRNN for an extended audio analysis and listening range.
    \item A multi-orientation causal video prediction layer built upon a novel Directional ConvGRU.
\end{itemize}




\section{Main architecture}

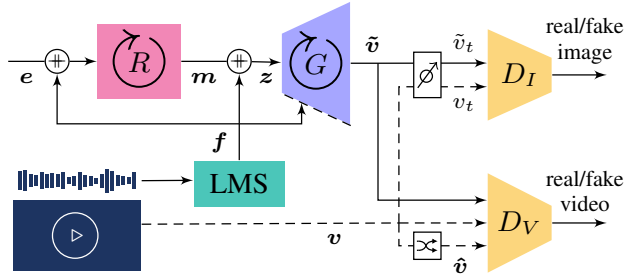
\begin{figure}[t]
    \centering
    \input{diagrams/avgan_arch}
    \caption{Main architecture. The training video is batched in consecutive $T$-frame sequences. Audio features $\mbs{f}$ (LMS) are routed to: (1) a recurrent neural network $R$ as source of motion concatenated with noise $\mbs{e}$, (2) the generator $G$ as source of content concatenated with motion tokens $\mbs{m}$, and (3) generative instance normalization layers. $R$ comprises a residual multi-scale DilatedRNN. $G$ contains Directional ConvRNNs. The video discriminator $D_V$ learns from reals $\mbs{v}$, fakes $\mbs{\tilde{v}}$, and shuffled versions $\mbs{\hat{v}}$. The image discriminator $D_I$ receives real $v_t$ and synthetic $\tilde{v}_t$ frames with the same random index.}
    \label{fig:avgan_arch}
\end{figure}

The proposed model is made of four main neural networks, as illustrated in \cref{fig:avgan_arch}. A recurrent neural network $R$ models the physically plausible motion paths over time. Next, a generator $G$ maps audio-guided motion tokens into video sequences, as close as possible to real distributions. Then, the realism of $G$'s outputs is compared to real training samples by a couple of discriminators. Thus, $D_I$ focuses on individual images, while $D_V$ on the notion of motion by criticizing video sequences.

\newcommand{\Tau}{\,\mathrm{T}}

\subsection{Sound representation}
\label{sec:sound_representation}
Extracting features from sound is a fundamental piece in the whole pipeline. 
Spectrogram representation is a common technique in single audio domain GANs \cite{Donahue2019AdversarialAS, Kumar2019MelGANGA, Nistal2020DrumGANSO} and is closely related to human hearing. 
In practice, the spectrogram is computed applying on a sliding window series the Short-Time Fourier Transform (STFT), to which the Log Mel Spectrogram (LMS) is a commonly applied perceptual non-linearity \cite{chen2017deep, hao2018cmcgan, duan2021cascade, Lee2022SoundGuidedSV}.

Previous music-to-image methods, above mentioned, use spectrograms in their original bidimensional shape through a symmetric UNet-like generator. However, such an approach is too rigid and unnecessarily symmetric to scale up towards higher resolutions.
On the other hand, the speech-to-video methods tend to draw upon more flexible representations.

Let $\Phi$ denote a spectrogram of $b\!\in\![1..B]$ frequency bands, split in chunks $c_t\!=\!\{\Phi[b,H\cdot(t-1)+a]\;|\,a\!\in\![1..A]\}$ from a $t$-index sliding window of $A$ audio frames (time bins) and stride $H$.
Then, the sound features of each audio chunk can be defined as 
$f_t=\frac{1}{A}\sum_{a}c_t$.
The waveform is effectively converted into $B\!=\!64$ log mel frequency bins (filtering frequencies $0.125$-$7.5\,\mathrm{kHz}$) with a $25\,\mathrm{ms}$ window length and $10\,\mathrm{ms}$ hop size.

The duration of the audio chunks, \ie the temporal resolution, is critical. Human perception is highly sensitive to periodic waveforms ---such as speech or music--- in short to medium timescales ($1$-$100\,\mathrm{ms}$) \cite{Engel2019GANSynthAN}. 
In speech recognition fine-grained intervals, such as $10$-$20\,\mathrm{ms}$, are often used \cite{Malik2021AutomaticSR}, while the average phoneme duration for instance in English might be around $100\,\mathrm{ms}$ \cite{Crystal1988TheDO}.
From a music perspective, $75$-$100\,\mathrm{ms}$ corresponds to sixteenth-notes (\musSixteenth) at $\musQuarter=150$-$200$ beats per minute, which is a reasonable upper bound. 
We finally set an intermediate value of $85\,\mathrm{ms}$ ($A=8$). The effective temporal resolution will eventually be constrained by the video frame rate, whose inverse is the hop size,
hereafter fixed to $50\,\mathrm{ms}$ ($H=5$) or $20\,\mathrm{fps}$.
Note that chunk overlapping ---in this case $37.5\%$--- already encourages temporal coherency. Herein, the first video frame is taken for each audio chunk.
 



\subsection{Audio temporal coherency}
\label{sec:audio_coherency}


A basic yet effective way of inducing temporal correlation is by means of RNNs, which also helps to generate videos of varying length even beyond the training length \cite{balaji2019conditional}. Indeed, an auxiliary generator has better efficiency than a 3D generator alone \cite{saito2017temporal}, which struggles to generate coherent sequences beyond the training length, as corroborated here in preliminary experiments.

Among the RNN family, Gated Recurrent Unit (GRU) and Long Short-Term Memory (LSTM) are used indistinctly. In fact both perform comparable for speech and music analysis \cite{Chung2014EmpiricalEO}. Here, the former is favored given its more efficient memory and computation requirements.

At this point, an important design choice resides in source of motion and sound routing, see \cref{fig:avgan_arch}. These types of routing are not either mutually exclusive or redundant, in fact their combination is beneficial, as shown later. Hereafter, let $\mbs{m}=R(\mbs{r})\in\mathbb{R}^{T\times M}$ denote a $T$-enrolled and compact motion encoding of size $M$ performed by an RNN.

\medskip\noindent\textbf{Sound as content}. A precursor MoCoGAN disentangles content from motion by feeding the generator with two independent noise vectors. One represents frames as points in the latent space and the other encodes a latent motion path between frames. Thus, $\mbs{r}:=\mbs{e}=\{e_t\in\mathbb{R}^E|t\in[1..T]\}$ is a $T$-length sequence of random vectors of normal distribution $\mathcal{N}(0,1)$, where typically $M\!=\!E$. Here, content noise is replaced by sound features $\mbs{f}=\{f_t\in\mathbb{R}^B|t\in[1..T]\}$, so that the generator's input is concatenated channel-wise across the temporal dimension as $\mbs{z}:=[\mbs{f},\mbs{m}]\in\mathbb{R}^{T\times Z}$, with $Z\!=\!B\!+\!E$. An important strength of this approach is that the generator has direct access to the raw sound features at each time step.

\medskip\noindent\textbf{Motion from sound}. In this approach the sound features are passed through the recurrent network, sometimes concatenated with motion noise as $\mbs{r}:=[\mbs{f}, \mbs{e}]$ of size $M=B+E$, so that the generator simply receives $\mbs{z}:=\mbs{m}$.
This is a preferred configuration of most audio-driven video generation methods \cite{Song2019TalkingFG, Vougioukas2019RealisticSF, Song2019TalkingFG, Chen2019HierarchicalCT, Li2020DirectST, balaji2019conditional, Lee2022SoundGuidedSV}.
Albeit it entails a much harder task, since motion associated with elementary sound events ---notes or phonemes--- and their transitions needs to be encoded by a simpler recurrent network.

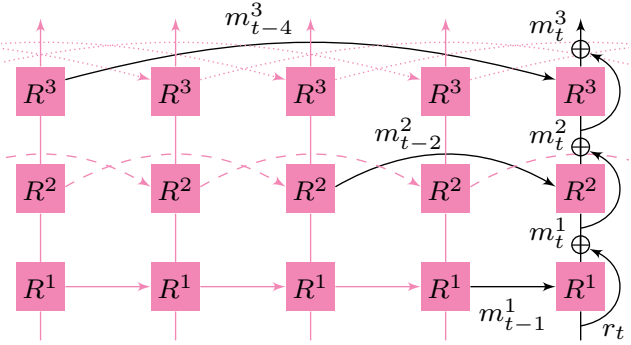
\begin{figure}[t]
    \centering
    \input{diagrams/dilated_rnn}
    \caption{Motion encoding: a 3-layer DilatedRNN with residual connections (depicted only for the current time step). Note recurrent cells have the same input-output size.}
    \label{fig:dilated_rnn}
\end{figure}

\medskip\noindent\textbf{Dilated and residual recurrency}. A multi-layer RNN with dilated skip connections is proposed to deal with more complex temporal dynamics of sound, which sometimes unfolds at different resolution speeds. Also, residual connections are routed to facilitate the propagation of audio features through the network, see \cref{fig:dilated_rnn}. To formalize that, let's reformulate the recurrent encoding as:

\begin{equation}
\label{eq:dilated_rnn}
    m_t^l = \varphi(R^l(m_t^{l-1}, m_{t-d}^l) + m_t^{l-1})
\end{equation}

where $m_t^l$ is the output at time step $t$ of the recurrent cell at layer $l\in[1..L_R]$, which receives the previous layer's output and the skipped hidden state with dilation factor $d=2^l$. $\varphi$ stands for an activation function, \eg LeakyReLU. For instance, with the settings in \cref{sec:sound_representation}, $L_R=3$ stacked layers result in $50$, $100$, and $200\,\mathrm{ms}$ audio hops.

\medskip\noindent\textbf{Single sequence batch size}.
A key architectural aspect herein falls to batch size and overfitting a single video. Most implementations reset the RNN's init state after each training iteration, either with zeroes or a random vector. By doing this, multiple video sequences can be batched in a single training pass. Indeed, large batches are beneficial also for GANs \cite{Brock2019LargeSG}. Nevertheless, we observed that the unrolling of temporally ordered batches of size 1 with consecutive video sequences was critical for convergence, so the last internal state becomes the initial state of the following batch. 



\subsection{The adversarial networks}
\label{sec:thenetworks}

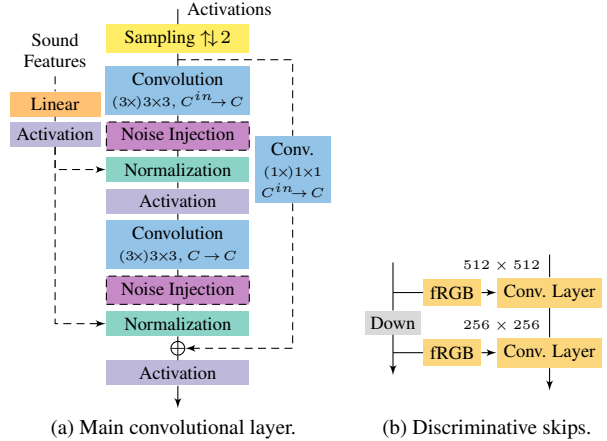
\begin{figure}[t]
    \centering
    \input{figures/building_blocks}
    \caption{Main building blocks made of a series of convolutional, normalization, and activation layers (LeakyReLU).
    The generator uses audio conditional instance normalization, noise injection, and residual connections.
    The discriminators instead use batch normalization, an extra temporal dimension (video), and skip connections (b) through convolutional layers (fRGB) transforming latent to color spaces.
    }
    \label{fig:building_blocks}
\end{figure}

The generator and both discriminators are built upon a series of feedforward layers, made of convolutional, normalization, and activation layers, see \cref{fig:building_blocks}. While the discriminators use batch normalization, the generator uses audio conditional instance normalization and noise injection \cite{feng2021understanding, karras2019style}.


Sampling layers before every convolutional layer performed best in this work.
Although most implementations prefer a kernel size 4 and sampling stride 2 \cite{isola2017image, Brock2019LargeSG, karras2019style}, a symmetric configuration with kernel size 3 and stride 1 was utilized. Either way, the output resolution doubles as the network progresses to the outermost layers. Nearest-neighbor interpolation is used for up-sampling, while 2D-3D average pooling is used for down-sampling. 

The number of channels doubles as each network progresses to the innermost layers, noted generically as $C$ for simplicity.
Convolutional layers with kernel size 1 are used to convert between a $C$-channel latent and 3-channel color spaces, supplemented with $tanh$ gates.

To scale-up the network towards high-resolution regimes, different configurations from \cite{karras2020analyzing} were tested in \cref{sec:model_assessment}, using residual and skip connections respectively for $G$ and $D$s.

\medskip\noindent\textbf{Audio conditional generator}.
A format layer first replicates both spatial dimensions to accommodate the input vector $\mbs{z}\in\mathbb{R}^{T\times Z}$ into the first generative layer, until reaching dimensionality $\mathbb{R}^{T\times Z\times4\times 4}$. Then a convolutional layer outputs $\mathbb{R}^{T\times C\times 4\times 4}$ latent vectors. Thus, the output video size $N\!\times\!N$ relates to the number of layers as $L=log_2(N)-2$. Note that a batch of 1 sequence is unrolled and the generator effectively receives $T$ frames per training iteration.

In order to reinforce access to conditional data traveling through the generator, a well-known strategy consists in embedding class vectors modulating normalization layers \cite{karras2019style, Brock2019LargeSG}, which is seemingly exploited also in conditional video generation \cite{tian2021a, skorokhodov2022stylegan, Lee2022SoundGuidedSV}, while others prefer disentangled generative layers \cite{wang2020g3an}. Here, as illustrated in \cref{fig:layer}, each sound feature vector $f_t$, intuitively related to a set of related visual poses or layouts, is encoded into a class feature vector, ultimately modulating a conditional instance normalization layer. This can be formalized for each generative layer as:

\begin{equation}
\begin{aligned}
    \gamma_t = \varphi(f_t\cdot W_\gamma + b_\gamma) \\
    \beta_t = \varphi(f_t\cdot W_\beta + b_\beta) \\
    \hat{x}_t = \gamma_t \left(\frac{x_t-\mu_t}{\sigma_t}\right) + \beta_t
\end{aligned}
\end{equation}

where $\mu_t$ and $\sigma_t$ are the mean and variance across spatial dimensions of the actual input activations $x_t$. The matrices $W_\gamma$, $W_\beta\in\mathbb{R}^{B\times C}$ are weights and $b_\gamma$, $b_\beta\in\mathbb{R}^C$ biases of each learnable linear transformation. $\varphi$ is an activation function, \eg LeakyReLU.


\medskip\noindent\textbf{Image and video discriminators}.
Both $D_I$ and $D_V$ have the same number of layers as $G$.
At their output a decision layer embeds $C$ channels into $1$ by means of $1\!\times\!1$ convolutions as PatchGAN to encourage discrimination in local patches and potentially improve high-frequency details \cite{isola2017image, Pumarola2018GANimationAF, tulyakov2018mocogan, duan2021cascade}.

$D_V$ is built on 3D kernels, one temporal and two spatial dimensions equally sampled at the beginning of each layer, see \cref{fig:layer}. Note the temporal dimension can be downsampled $\lfloor log_2(T)\rfloor$ times, \eg  $T=32$ sequences allow for temporal reduction only along the first $5$ layers.

For each forward pass $D_I$ receives a random frame $\tilde{v}_t$ from a synthetic video sequence $\mbs{\tilde{v}}$. The same random index is used to pick up a real frame $v_t$ from a source sequence $\mbs{v}$. Presumably, by sharing the same index, the mapping between sound features and visual attributes is facilitated. $D_V$ receives real $\mbs{v}$ and fake $\mbs{\tilde{v}}$ video sequences, as well as a shuffled version $\mbs{\hat{v}}$ to reinforce temporal coherency \cite{balaji2019conditional}.



\subsection{Video temporal coherency}
\label{sec:video_coherency}

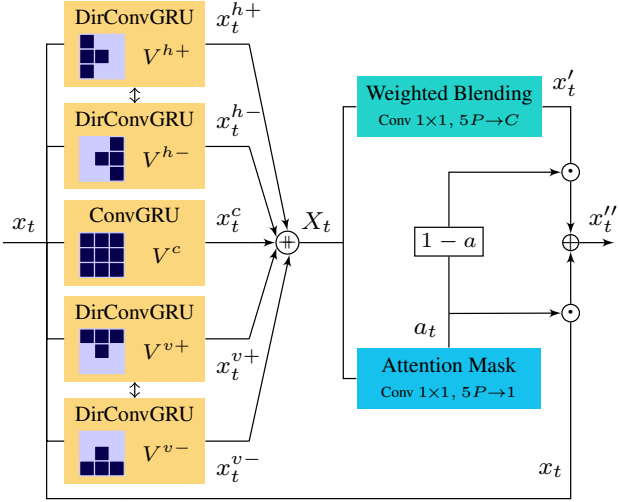
\begin{figure}[t]
    \centering
    \input{diagrams/video_prediction.tex}
    \caption{Video prediction layer: 4-directional and 1-centered ConvGRUs (kernel size 3). Spatial predictions are channel-wise concatenated into $X_t$ and blended by $1\!\times\!1$ convolutions to accommodate output channels. Merged predictions $x'_t$ and previous hallucinated activations $x_t$ contribute to the output $x''_t$ according to an auto-regressive mask $a_t$ shared across channels with the Hadamard product $\odot$. Note opposite directions share weights.}
    \label{fig:video_prediction}
\end{figure}

A novel recurrent multi-directional convolutional layout is proposed for causal video prediction. The activation distribution at time step $t$ can be expressed as $x_t^d\sim p(x_t|x_{t-1}^d)$, where $x$ are the hallucinated output activations of a generative layer and $x^d$ directional predictions. To do so, a video prediction layer is formed by 4-directional ConvGRUs (DirConvGRU), predicting positive and negative motion in horizontal and vertical directions, as well as a squared-centered ConvGRU, dealing with motion along the camera axis. See \cref{fig:video_prediction}.

An amenable implementation is built under the assumption that opposite directions can share weights \cite{Byeon2018ContextVPFC} and directionality can be achieved by aggregating increasingly larger and shifted one-dimensional kernels.

Let
$h^{+}[i,j]=\left\{\begin{matrix}
w\in \mathbb{R} & |i|\!<\!K,\,j=1\\ 
0 & elsewhere\\
\end{matrix}\right.$ 
be a $1$-shifted $K\!\times\!1$ kernel predicting negative horizontal movements, and $\tau_h^{-}$ a negative horizontal translation operator, such that $(\tau_h^{-}h)[i,j]=h[i,j+1]$. By the translational equivariance property of convolutions it is straightforward that $x*h^{+}=\tau_h^{+}(\tau_h^{-}(x*h^{+}))=\tau_h^{+}(x*\tau_h^{-}h^{+})=\tau_h^{+}x*h$, where $h[i,j]=h^+[i,j+1]$ is the centered version of $h^{+}$. Similar reasoning can be done for the equivalent horizontal $h^{-}$ and vertical $v^{+}$, $v^{-}$ kernels. Thus, directional kernels of triangular shape inside DirConvGRUs are now easily constructed by adding up activations of multiple $k$-shifted kernels of increasing size $K=2k+1$ and shared bias. Furthermore, this allows to implement weight sharing between opposite directions by utilizing the same DirConvGRU layer, whose kernels' outputs are properly shifted according to the desired direction.

The video prediction layer is stacked after the outermost generative layer, although it can be potentially inserted in each of the generative layers.
The video prediction activations are finally concatenated as:
\begin{equation}
    \begin{aligned}
    X_t=[x^{h+}_t,x^{h-}_t, x^{c}_t, x^{v+}_t, x^{v-}_t]
    \end{aligned}
\end{equation}

where $x^d_t=V^d(x_t,x^d_{t-1})\in\mathbb{R}^{P\times N\times N}$ is the $t$-th hidden state, also output, of a $K$-size centered or directional ConvGRU for $d\in\{h+,h-,c,v+,v-\}$.

The next step deploys a combination strategy. A weighted blending layer not only merges all directional predictions \cite{Byeon2018ContextVPFC} but also predicts an auto-regressed attention mask \cite{Pumarola2018GANimationAF, Chen2019HierarchicalCT, wang2018video, Chan2019EverybodyDN}, which helps to merge both the hallucinated $x_t$ and predicted $x'_t$ pixels as follows:

\begin{equation}
\begin{aligned}
    x'_t=\varphi(X_t^\intercal\cdot W_x + b_x)\\
    a_t=\sigma(X_t^\intercal\cdot W_a + b_a) \\
    x''_t = a_t \odot x_t + (1 - a_t) \odot x'_t
\end{aligned}
\end{equation}

where the blending $W_x\in \mathbb{R}^{(5\times P)\times C\times N\times N}$ and masking weights $W_a\in \mathbb{R}^{(5\times P)\times 1\times N\times N}$ are implemented as $P$-channel $1\!\times\!1$ convolutions ($P\!=\!C$ for convenience). The activation functions $\sigma$ and $\varphi$ are sigmoid and the identity respectively ---LeakyReLU performed slightly worse. Also, $\odot$ denotes the channel-wise Hadamard product, and $x_t''$ is the hallucinated output with induced temporal coherency.


\section{Training and loss function}

The importance of the loss function has sometimes been questioned \cite{Lucic2018AreGC, kurach2019large}. Nevertheless, no other apart from Wasserstein with gradient penalty (WGAN-GP) \cite{Gulrajani2017ImprovedTO} was able to provide training stability and image quality in this work.
In addition, perceptual loss \cite{Johnson2016PerceptualLF} has shown a remarkable capability to synthesize sharper details \cite{Song2019TalkingFG, Chen2018LipMG, wang2018video, Wang2018HighResolutionIS, Chan2019EverybodyDN, zhu2021let, tian2021a}. In a preliminary study, it obtained about $5$dB better accuracy, about $\times 2.5$ faster convergence, and notably more stable training compared to the $L_1$-norm.

Let $\mbs{v}\!=\!\{v_t\in\mathbb{R}^{3\times N\times N}|\,t\in[1..T]\}$
represent a ground truth video sequence, paired for each audio chunk $c_t$ with its sound representation $\mbs{z}\!=\!\{z_t\in\mathbb{R}^Z|\,t\in[1..T]\}$, as described in \cref{sec:sound_representation}. Likewise, a synthesized video sequence is implicitly defined as $\mbs{\tilde{v}}=G(\mbs{z})$.
Then, the minimax adversarial loss function, expressed in two terms $\displaystyle \min_{R,G}\max_{D_I,D_V}=\mathcal{L}_I+\mathcal{L}_V$, is defined as:

\begin{equation}
\begin{aligned}
    \mathcal{L}_I=\!\mathop{\mathbb{E}}_{v_t\sim \mathbb{P}_{v_t}}\!\!\![D_I(v_t)]-\!\!
    \mathop{\mathbb{E}}_{\tilde{v}_t\sim \mathbb{P}_{\tilde{v}_t}}\!\!\![D_I(\tilde{v}_t)]
    + \lambda\;\mathcal{P}_{D_I}(v_t,\tilde{v}_t)
\end{aligned}
\end{equation}   

\begin{equation}
\begin{aligned}
    \mathcal{L}_V=\mathop{\mathbb{E}}_{\mbs{v}\sim \mathbb{P}_v}\!\![D_V(\mbs{v})]-
    \mathop{\mathbb{E}}_{\mbs{\tilde{v}}\sim \mathbb{P}_{\tilde{v}}}\!\![D_V(\mbs{\tilde{v}})]
    -\mathop{\mathbb{E}}_{\mbs{\hat{v}}\sim \mathbb{P}_{\hat{v}}}\!\![D_V(\mbs{\hat{v}})]\\
    +\;\lambda\;\mathcal{P}_{D_V}(\mbs{v},\mbs{\tilde{v}})+\lambda\;\mathcal{P}_{D_V}(\mbs{v},\mbs{\hat{v}}) + \alpha\, ||\phi(\mbs{v})-\phi(\tilde{\mbs{v}})||_2
\end{aligned}
\end{equation}

where each discriminator's gradient penalty is expressed as $\mathcal{P_D}(x,y)\!=\!\mathop{\mathbb{E}}_{x,y\sim \mathbb{P}_{x,y}}[(||\nabla_{x,y}D(x\cdot\epsilon+y\cdot(1\!-\epsilon))||_2-1)^2]$, related to source $\mathbb{P}_v$, modeled $\mathbb{P}_{\tilde{v}}$, and shuffled $\mathbb{P}_{\hat{v}}$ video data distributions. Individual image distributions are $t$ subscripted. The perceptual loss uses $\phi$ to denote the output features of an intermediate VGG layer. Hyper-parameters $\alpha$ and $\lambda$ are meant to control the importance of loss terms during training.
For every training iteration, alternating gradient updates are conducted. First $D_I$ and $D_V$ are updated while fixing $G$ and $R$, and vice versa.


\section{Experiments}
\label{sec:experiments}

Training on a single video might give the impression of being an easy-to-learn task. However, other normalization types and loss functions led to divergence or model collapse. Moreover, successful setups needed about a training day ($50\mathrm{k}$ iterations) for $128\!\times\!128$ and about 3 days ($100\mathrm{k}$ iterations) for $256\!\times\!256$ to reach a decent image quality\footnote{Intel(R) Core(TM) i9-10900X CPU @ 3.70GHz and NVIDIA GeForce RTX 3090.}. Full-model higher resolutions required far beyond $24\mathrm{GB}$ GPU-memory and too long training runs that were finally discarded.



\subsection{Hyper-parameters setup}

Batched video sequences have $L=32$ frames forwarded one at a time.
The size of random noise vectors is $E=2$ (just a guess).
The input and hidden recurrent layers have the same size $M=66$,
so eventually the generator ingests vectors of size $Z=130$.
Image resolution $256\!\times\!256$ is achieved with $6$ generative layers. The outermost one has $C=16$ channels, incremented in powers of two, with $512$ maximum channels.
Images are $[-1, 1]$ normalized.
$R$ uses normal initialization and zeroed init states (no prior).
Convolutional weights receive a random initialization from a normal distribution $\mathcal{N}(0,0.02)$, while DirConvGRUs use orthogonal initialization. All LeakyReLUs have $0.2$ negative slope.
TTUR learning rates are heuristically set to $10^{-4}$ for $G$ and $R$, while $D_I$ and $D_V$ updates at $4\cdot 10^{-4}$, without scheduling. The Adam optimizers \cite{Kingma2015AdamAM} have momentums of $0.3$ and $0.999$, with $\alpha\!=\lambda\!=\!10$.

\subsection{Model assessment}
\label{sec:model_assessment}

The goal of this work is to improve GANs ability to regenerate videos based exclusively on its audio as guidance, as realistic and coherent as possible, to later be able to re-animate the same visual content in synchrony with a replaced audio input. Text- or pose-based audio-to-video methods are not directly comparable within our aim. Furthermore, this study entails a particular one-shoot training on a single video. Therefore, a baseline implementation, with a common MfS scheme and the state of the art advances here described, is used to compare with each of the features proposed in this work.


\begin{figure}[h]
    \centering
    \includegraphics[width=0.32\linewidth]{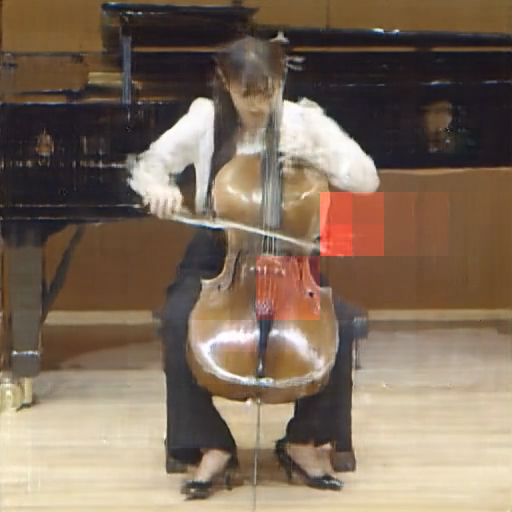}
    \includegraphics[width=0.32\linewidth]{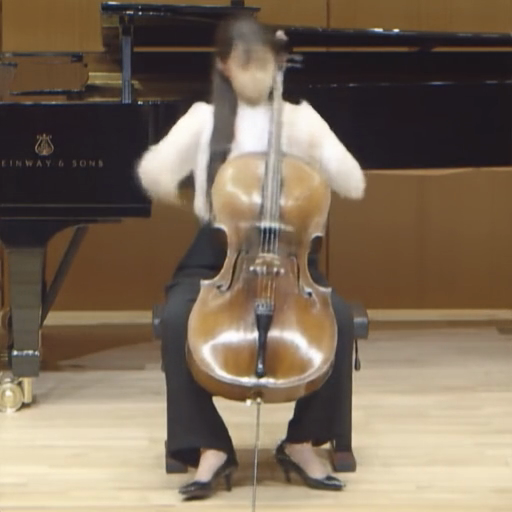}
    \includegraphics[width=0.32\linewidth]{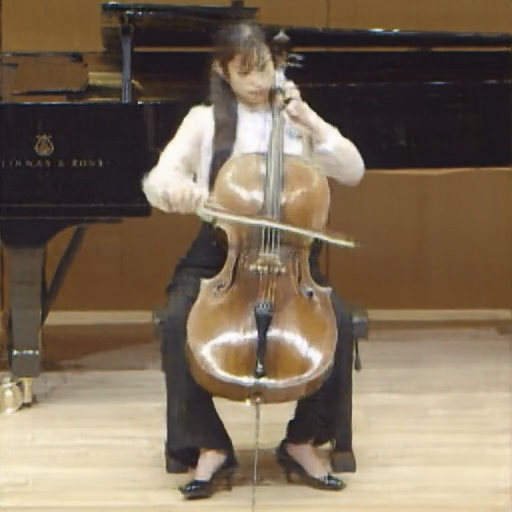}
    \caption{Illustration of artifacts produced by a $512\!\times\!512$ vanilla GAN with 
    (left) skip-connections, (middle) residual connections, and (right) a residual generator and skip-connected discriminators.}
   \label{fig:arc_artifacts}
\end{figure}

\begin{table*}[h!t]
    \centering
    \input{tables/ablation.tex}
    \caption[Ablation study]{Ablation study at $128\!\times\!128$ resolution
    on diverse audiovisual content, namely
    a cello\protect\footnotemark (melodic),
    a classic quintet\protect\footnotemark (harmonic),
    drums\protect\footnotemark (percussive),
    and a side-view talking head\protect\footnotemark (speech).
    Taking motion from sound (MfS) as a baseline, the sound routing features are activated one at a time: Residual+Dilated-RNN, sound as content (SaC), and audio conditional instance normalization (acIN). As for the video prediction group, all routing features are activated to compare between a basic ConvGRU and the proposed DirConvGRUs ($K\!=\!3$). Averages calculated over batches of the 20\%-split source video.}
    \label{tab:ablation}
\end{table*}

\addtocounter{footnote}{-3}
\footnotetext[\thefootnote]{Cello: \tiny\url{https://www.youtube.com/watch?v=zah9B0toTBQ}}
\addtocounter{footnote}{1}
\footnotetext[\thefootnote]{Quintet: \tiny\url{https://www.youtube.com/watch?v=R-Tk7-Ytes4}}
\addtocounter{footnote}{1}
\footnotetext[\thefootnote]{Drums: \tiny\url{https://www.youtube.com/watch?v=7H4hrsb_0tE&t=59s}}
\addtocounter{footnote}{1}
\footnotetext[\thefootnote]{Talking head: MEAD M31 \cite{wang2020mead}.}


\medskip\noindent\textbf{Skip vs. residual connections}.
In contrast to the observed in \cite{karras2020analyzing}, the layer connection has a strong impact on the synthesis quality, where the main architectural difference with this work is the spatio-temporal cross-domain generation. As illustrated in \cref{fig:arc_artifacts}, skip connections in the generator tend to produce large colored patches, increasing their presence and intensity as the input audio deviates from the original audio. Such glitches are probably induced by incorrect decisions taken in coarse low-level layers, and freely propagated through the outermost and finest layers. By switching to residual connections the artifacts simply disappear.
Conversely, discriminators built upon residual connections often produce severe blurring and erase some details, see the face and bow in \cref{fig:arc_artifacts} (middle). Instead, a residual generator and skip-connected discriminators usually produce sharper images, see \cref{fig:arc_artifacts} (right). The fact that finer details can travel more cleanly to the low-level layers might help the image discriminator fit more precise data distribution and indirectly force the generator to synthesize more realistic images. The progressive growth generally lead to an unstable training, specially after each scale activation, and the multi-scale gradients variant was unable to conform coherent imaging.


\medskip\noindent\textbf{Ablation study}.
A series of experiments were conducted on diverse audiovisual content of various minutes in length, as summarized in \cref{tab:ablation}.
Two perceptual objective image quality metrics, SSIM \cite{Wang2004ImageQA} and LPIPS \cite{Zhang2018TheUE}, and one video quality metric FVD\footnote{From this implementation \cite{skorokhodov2022stylegan}.} \cite{Unterthiner2018TowardsAG} were used.  We observed a clear tendency to objectively improve not only image but also video quality by adding independently each proposed feature. Moreover, video prediction layers, and in particular DirConvGRU more than a basic ConvGRU, provided an extra boost of temporal enhancement.

\medskip\noindent\textbf{Robustness to sound replacement}.
Synthesized images get affected when the distribution of feedforward sound deviates from the training source. This could be of interest to re-animate a video by using sounds from different contexts or acoustics.
To evaluate this, random audio clips were selected from FSD50K \cite{fonseca2022FSD50K}.
Since large variations in distribution are plausibly expected, specially in terms of temporal dynamics, we cannot trust on FVD, while SSIM and LPIPS are simply infeasible in the absence of ground-truth pairs. Instead, FID \cite{Heusel2017GANsTB} can measure distortions by comparing how far real and synthetic images are at high-level representation. From the results in \cref{tab:FSD50K}, the full model proves more robustness on in-the-wild sound replacement when trained on a music video, while apparently the generalization capability reduces when trained on speech.

\begin{table}
    \centering
    \resizebox{0.8\linewidth}{!}{\input{tables/base_vs_full_FSD50K.tex}}
    \caption{Baseline (MfS) and full model sound robustness comparison for the videos in \cref{tab:ablation}. FID averaged over $500$ audio clips of $1$-$5\,\mathrm{s}$ randomly selected from FSD50K \cite{fonseca2022FSD50K}.
    }
    \label{tab:FSD50K}
\end{table}

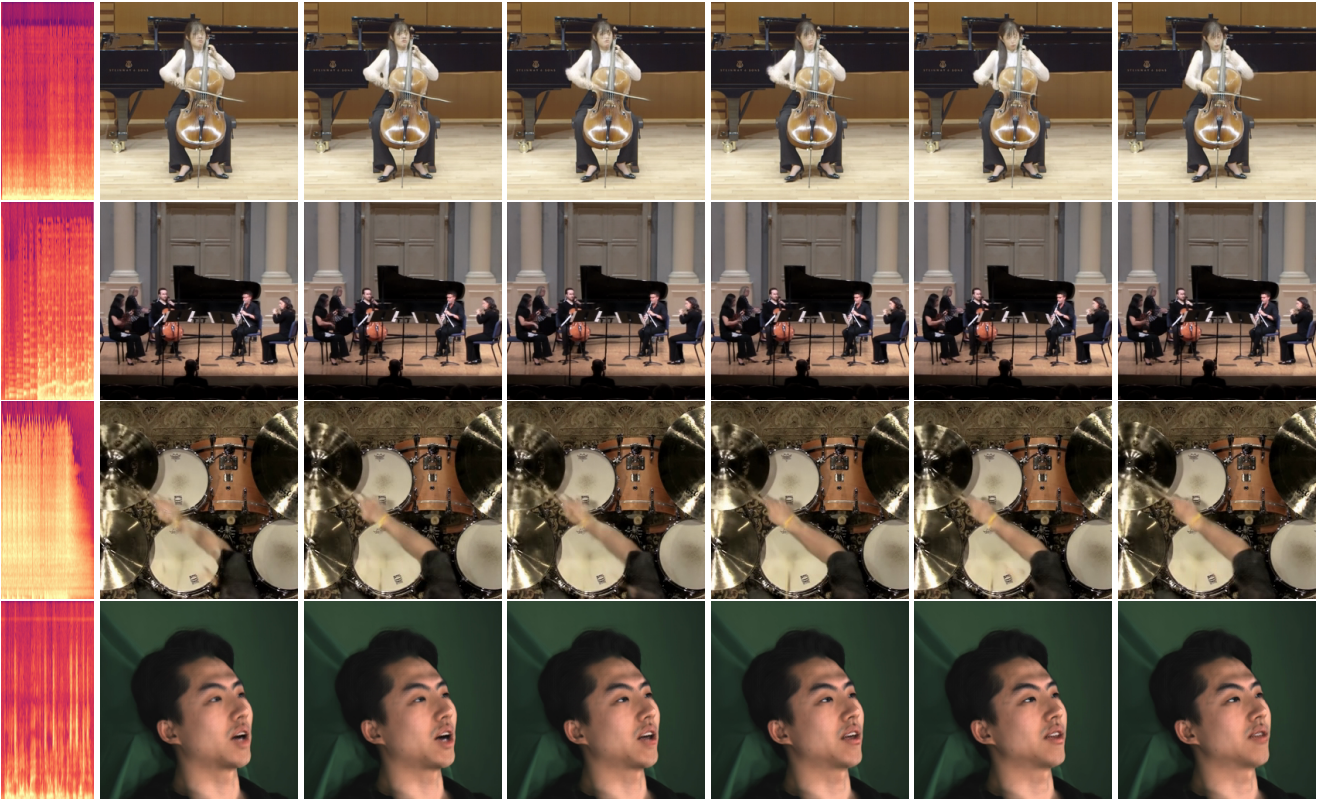
\begin{figure*}[t]
    \centering
    \input{figures/synth_seq_results}
    \caption{
    Examples of consecutive frames synthesized by our full-model at $256\!\times\!256$ resolution and $20\mathrm{fps}$ fed with validation audio samples of the videos in \cref{tab:ablation}.
    }
   \label{fig:video_seq_examples}
\end{figure*}

\medskip\noindent\textbf{Qualitative results}.
Our model is able to synthesize expressive long video sequences by synchronously responding to audio, without apparent degradation over time. Some illustrative examples are shown in \cref{fig:video_seq_examples}. The combined sound routing significantly reduces artifacts, especially when the input sound deviates from the original audio distribution, in accordance with \cref{tab:ablation} and \cref{tab:FSD50K}. Also, the video prediction layer tends to generate smoother motions and reduced flickering. Nevertheless, the persistence of temporal artifacts certainly affects the overall realism, specially for complex or sound-uncorrelated visual dynamics.

\section{Discussion}
\label{sec:discussion}

In this work various strategies to improve temporal coherence of sound-to-video GANs have been proposed, which can be applied straightforwardly to other domains. Many other configurations were tested, including sound encoding \cite{Owens2018AudioVisualSA, Wan2019TowardsAT}, conditioning augmentation \cite{zhang2017stackgan}, spectral and pixel normalization, joint audio-image discriminators, or feature matching loss \cite{wang2018video, Chan2019EverybodyDN}. Most of them did not show improvements and, in general, any strategy to encode (compress) audio features led to an impairment.

\medskip\noindent\textbf{Limitations}. While our method struggles to reproduce large and complex motion dynamics, specially those unrelated to sound, it is particularly demanding on both computation and memory despite our amenable DirConvRNN.

\medskip\noindent\textbf{Broader impact}. Generative models like this can certainly leverage smart tools for audiovisual content creation. On the other hand, unlike images, there is still a significant gap to achieve high-quality realistic videos. Even so, it may already raise important ethical concerns on inappropriate usage, from faking reality to illicit activities.

\section*{Acknowledgements}
\noindent This work was financially supported by the Catalan Government through the funding grant ACCIÓ-Eurecat (Project PRIV - DeepArts).

{\small
\bibliographystyle{ieee_fullname}
\bibliography{egbib}
}

\end{document}

%% file: diagrams/avgan_arch.tex
\tikzstyle{styrnn}=[fill=\rnncolor, font=\small, minimum size=2em]
\tikzstyle{stygen}=[fill=\gencolor, font=\small, minimum size=2em, rotate=90, inner sep=1pt, trapezium, trapezium stretches body, trapezium angle=65]
\tikzstyle{stydis}=[fill=\discolor, font=\small, minimum size=2em, rotate=-90, inner sep=1.8pt, trapezium, trapezium stretches body, trapezium angle=60]

\tikzstyle{stygenpath}=[]
\tikzstyle{stydispath}=[densely dashed]
\tikzstyle{styrnnpath}=[]

\tikzstyle{stysum}=[circle, draw=black, fill=white, minimum size=0mm, font=\tiny, inner sep=0pt]
\tikzstyle{styvid}=[fill=gray!30, text=white, minimum width=14mm, minimum height=7.8mm, fill={rgb,255: red,29; green, 52; blue,97}]
\tikzstyle{styaudio}=[minimum width=14mm, minimum height=4mm]
\tikzstyle{stystft}=[fill=\lmscolor, minimum width=10mm, minimum height=3mm, inner sep=4pt]
\tikzstyle{styopt}=[draw=black, fill=white, font=\tiny, inner sep=-1pt]

\newcommand{\randompick}[1][1]{%
  \begin{tikzpicture}[]
    \draw[-to] (#1*-1em,#1*-3ex) -- (#1*1em,#1*3ex);
    \draw (#1*-0.1em,#1*-0.3ex) circle (#1*1.4ex);
  \end{tikzpicture}%
}
\newcommand{\shuffle}[1][1]{%
  \begin{tikzpicture}[]
    \draw[-]
        (0.0mm,0.0mm) -- (0.2mm,0.0mm)
        (0.0mm,1.0mm) -- (0.2mm,1.0mm);
    \draw[-{Latex[length=0.6mm,width=0.6mm]}]
        (0.2mm,0.0mm) to [out=0,in=250] (1.0mm,0.5mm) to [out=70, in=180] (2.3mm,1.0mm);
    \draw[-{Latex[length=0.6mm,width=0.6mm]}]
        (0.2mm,1.0mm) to [out=0,in=110] (1.0mm,0.5mm) to [out=290, in=180] (2.3mm,0.0mm);        
  \end{tikzpicture}%
}

\begin{adjustbox}{width=\linewidth}
\begin{tikzpicture}[font=\scriptsize, auto,>=latex']
    
    \node [styrnn] (rnn) {\bottominset{$R$}{\Huge$\circlearrowright$}{5pt}{}};
    
    \node [stysum, left=3mm of rnn] (sum1) {$\mdoubleplus$};
    \coordinate [left=4mm of sum1] (input);
    \node (sum2) [stysum, right=5mm of rnn] {$\mdoubleplus$};

    \node [stygen] (gen) [below right=2.7mm and 4mm of sum2] {\rotatebox{-90}{\bottominset{$G$}{\Huge$\circlearrowright$}{5pt}{}}};

    \path[-] (input) edge node [below] {$\mbs{e}$} (sum1);
    \path[->] (sum1) edge (rnn);
        
    \path[-] (rnn) edge node [below] {$\mbs{m}$} (sum2);
    \path[->] (sum2) edge node [below] {$\mbs{z}$}(gen);
    
    \coordinate [right=3mm of gen.south] (genout);
    \draw[-,stygenpath] (gen.south) -- (genout) node [above, pos=0.8] {$\mbs{\tilde{v}}$};

    \node (video) [styvid, below left=15mm and 23mm of gen.south] {$\Large\triangleright$};
    \node (play) [circle, draw=white, minimum size=0.5cm, left=-0.94cm of video.west] {};    

    \node (audio) [styaudio, above=0mm of video] {};
    \foreach \x in {0,...,21}
        \node [, fill={rgb,255: red,29; green, 52; blue,97}, minimum width=0.04cm, inner sep=0pt, minimum height=(8*random()), left=-(0.06*\x+0.1) of audio.west] {};
    
    \node (stft) [stystft, below=9.5mm of sum2] {\small LMS};
    \coordinate[above=4mm of stft.north] (stftout);
    \draw[->] (audio.east) -- (stft.176);
    \draw[-] (stft.north) -- (stftout) node [left, pos=0.5] {$\mbs{f}$};
    \draw[->] (stftout) -| (sum1);
    \draw[->] (stftout) -- (sum2);
    \draw[->] (stftout) -| (gen.160);
    \draw[-, densely dashed] (gen.north west) -- (gen.bottom left corner);
    
    \node [stydis] (Dv) [above right=-14mm and 12mm of genout] {\rotatebox{90}{$D_V$}};
    \coordinate[left=9.7mm of Dv.south] (Dvin);
    \draw[-,stydispath] (video.10) -- (Dvin) node [below, pos=0.75] {$\mbs{v}$};
    \draw[->,stygenpath] (genout) |- (Dv.235);
    \draw[->,stydispath] (Dvin) -- (Dv.270);
    \draw[->,stydispath] (Dvin) |- (Dv.305) node [below, pos=0.85] {$\mbs{\hat{v}}$};
    \coordinate[right=6mm of Dv.north] (Dvout);
    \draw[->] (Dv.north) -- (Dvout) node [above, pos=0.6, align=center, font=\scriptsize] {real/fake\\video};

    \node [stydis] (Di) [above right=2.3mm and 12mm of genout] {\rotatebox{90}{$D_I$}};
    \draw[->,stygenpath] (genout) -- (Di.250) node [above left] {$\tilde{v}_t$};
    \draw[->,stydispath] (Dvin) |- (Di.290) node [below, pos=0.87] {$v_t$};;
    \coordinate[right=6mm of Di.north] (Diout);
    \draw[->] (Di.north) -- (Diout) node [above, pos=0.6, align=center, font=\scriptsize] {real/fake\\image};
    
    \node (piker) [styopt, minimum height=5mm, minimum width=3mm, left=5mm of Di.270] {\randompick[0.6]};

    \node (shuffle) [styopt, minimum size=3mm, left=5mm of Dv.305] {\shuffle[1]};

\end{tikzpicture}
\end{adjustbox}

%% file: diagrams/dilated_rnn.tex
\tikzstyle{sty}=[minimum width=5mm, minimum height=5mm, inner sep=0.5pt]

\tikzstyle{foresty}=[black]
\tikzstyle{backsty1}=[\rnncolor,dashed]
\tikzstyle{backsty2}=[\rnncolor,densely dotted]
\tikzstyle{backsty3}=[\rnncolor]

\begin{adjustbox}{width=\linewidth}
\begin{tikzpicture}[font=\footnotesize, node distance=14mm, auto, >=latex']

    \clip (-6.0,-0.6) rectangle (0.5,3.1);

    \node [sty,fill=\rnncolor] (cell00) {$R^1$}; 
    \node [sty,fill=\rnncolor] (cell10) [left of=cell00] {$R^1$};
    \node [sty,fill=\rnncolor] (cell20) [left of=cell10] {$R^1$};
    \node [sty,fill=\rnncolor] (cell30) [left of=cell20] {$R^1$};
    \node [sty,fill=\rnncolor] (cell40) [left of=cell30] {$R^1$};

    \path[->,foresty]
        (cell10) edge node [below] {$m_{t-1}^1$} (cell00);
        
    \path[->,backsty3]
        (cell20) edge (cell10)
        (cell30) edge (cell20)
        (cell40) edge (cell30);

    \node [stysum] (sum00) [above=0.8mm of cell00.north] {$+$};

    \node [sty,fill=\rnncolor] (cell01) [above=5mm of cell00, align=center] {$R^2$};
    \node [sty,fill=\rnncolor] (cell11) [above=5mm of cell10, align=center] {$R^2$};
    \node [sty,fill=\rnncolor] (cell21) [above=5mm of cell20, align=center] {$R^2$};
    \node [sty,fill=\rnncolor] (cell31) [above=5mm of cell30, align=center] {$R^2$};
    \node [sty,fill=\rnncolor] (cell41) [above=5mm of cell40, align=center] {$R^2$};

    \path[->,foresty] (cell21.5) edge[bend left=30] node [above left=-1mm and -1mm] {$m_{t-2}^2$} (cell01.175);
    \path[->,backsty1]
        (cell31.5) edge[bend left=30] (cell11.175)
        (cell41.5) edge[bend left=30] (cell21.175);

    \node [sty] (cell51) [left of=cell41, align=center] {};
    \node [sty] (rell11) [right of=cell01, align=center] {};

    \path[->, backsty1]
        (cell51.5) edge[bend left=30] (cell31.175)
        (cell11.5) edge[bend left=30] (rell11.175);

    \path[-,foresty]
        (cell00)  edge node [left, pos=0.6] {$m_t^1$} (cell01);
        
    \path[-,backsty3]
        (cell10)  edge (cell11)
        (cell20)  edge (cell21)
        (cell30)  edge (cell31)
        (cell40)  edge (cell41);

    \node [stysum] (sum01) [above=0.8mm of cell01.north] {$+$};
    
    \node [sty,fill=\rnncolor] (cell02) [above=5mm of cell01, align=center] {$R^3$};
    \node [sty,fill=\rnncolor] (cell12) [above=5mm of cell11, align=center] {$R^3$};
    \node [sty,fill=\rnncolor] (cell22) [above=5mm of cell21, align=center] {$R^3$};
    \node [sty,fill=\rnncolor] (cell32) [above=5mm of cell31, align=center] {$R^3$};
    \node [sty,fill=\rnncolor] (cell42) [above=5mm of cell41, align=center] {$R^3$};

    \path[->,foresty] (cell42.25) edge[bend left=15] node [above left=-0.6mm and 0.5mm] {$m_{t-4}^3$} (cell02.155);

    \node [sty] (cell52) [left of=cell42, align=center] {};
    \node [sty] (cell62) [left of=cell52, align=center] {};
    \node [sty] (cell72) [left of=cell62, align=center] {};

    \node [sty] (rell12) [right of=cell02, align=center] {};
    \node [sty] (rell22) [right of=rell12, align=center] {};
    \node [sty] (rell32) [right of=rell22, align=center] {};
    
    \path[->, backsty2]
        (cell12.25) edge[bend left=15] (rell32.155)
        (cell22.25) edge[bend left=15] (rell22.155)
        (cell32.25) edge[bend left=15] (rell12.155)
        (cell52.25) edge[bend left=15] (cell12.155)
        (cell62.25) edge[bend left=15] (cell22.155)
        (cell72.25) edge[bend left=15] (cell32.155);

    \path[-,foresty]
        (cell01)  edge node [left, pos=0.6] {$m_t^2$} (cell02);
        
    \path[-,backsty3]
        (cell11)  edge (cell12)
        (cell21)  edge (cell22)
        (cell31)  edge (cell32)
        (cell41)  edge (cell42);

    \node [stysum] (sum02) [above=0.8mm of cell02.north] {$+$};
    
    \coordinate[below=3mm of cell00] (cell0in);
    \coordinate[below=3mm of cell10] (cell1in);
    \coordinate[below=3mm of cell20] (cell2in);
    \coordinate[below=3mm of cell30] (cell3in);
    \coordinate[below=3mm of cell40] (cell4in);

    \path[-,foresty]
        (cell0in)  edge node [right=1.0mm, pos=0.3] {$r_t$} (cell00);
    \path[-,backsty3]
        (cell1in)  edge (cell10)
        (cell2in)  edge (cell20)
        (cell3in)  edge (cell30)
        (cell4in)  edge (cell40);

    \coordinate[above=5mm of cell02] (cell0out);
    \coordinate[above=5mm of cell12] (cell1out);
    \coordinate[above=5mm of cell22] (cell2out);
    \coordinate[above=5mm of cell32] (cell3out);
    \coordinate[above=5mm of cell42] (cell4out);

    \path[->,foresty] 
        (cell02)    edge node [above left=-1.1mm and 0.0mm] {$m_t^3$} (cell0out);
    \path[->,backsty3]        
        (cell12)    edge (cell1out)
        (cell22)    edge (cell2out)
        (cell32)    edge (cell3out)
        (cell42)    edge (cell4out);

    \coordinate[below=1.5mm of cell00.south] (res0);
    \coordinate[below=1.5mm of cell01.south] (res1);
    \coordinate[below=1.5mm of cell02.south] (res2);
    \coordinate[above=1.9mm of cell02.south] (res3);
    
    \draw[->] (res0) to [out=10,in=270] ($(cell00.east)+(1.5mm,0)$) to [out=90, in=-25] (sum00);
    \draw[->] (res1) to [out=10,in=270] ($(cell01.east)+(1.5mm,0)$) to [out=90, in=-25] (sum01);
    \draw[->] (res2) to [out=10,in=270] ($(cell02.east)+(1.5mm,0)$) to [out=90, in=-25] (sum02);
        
\end{tikzpicture}
\end{adjustbox}

%% file: figures/building_blocks.tex
    \begin{subfigure}[b]{0.59\linewidth}
        \centering
        \input{diagrams/layer}
        \subcaption{Main convolutional layer.}
        \label{fig:layer}
    \end{subfigure}
    \begin{subfigure}[b]{0.39\linewidth}    
        \centering
        \input{diagrams/disc_layers}
        \subcaption{Discriminative skips.}
        \label{fig:skips_disc}
    \end{subfigure}

%% file: diagrams/layer.tex
\tikzstyle{sty}=[align=center, minimum width=19mm, inner sep=2.2pt]

\definecolor{sampcolor}{HTML}{ffee65}
\definecolor{convcolor}{HTML}{93c3e7}
\definecolor{noisecolor}{HTML}{c68ac7}
\definecolor{normcolor}{HTML}{8bd3c7}
\definecolor{relucolor}{HTML}{beb9db}
\definecolor{soundcolor}{HTML}{ffc174}
\definecolor{soundrelucolor}{HTML}{beb9db}

\begin{tikzpicture}[font=\scriptsize, auto,>=latex']
    \node [sty,fill=sampcolor] (sampling) {Sampling $\uparrow\!\downarrow\!2$};
    
    \node [sty,fill=convcolor] (conv1) [below=1.6mm of sampling] {Convolution\\[0mm]\tiny $(3\!\times\!)3\!\!\times\!\!3, C^{in}\!\!\!\rightarrow\!C$};

    \node [sty,fill=noisecolor, draw, densely dashed] (noise1) [below=0.8mm of conv1] {Noise Injection};
    \node [sty,fill=normcolor] (norm1) [below=0.8mm of noise1] {Normalization};
    \node [sty,fill=relucolor] (relu1) [below=0.8mm of norm1] {Activation};

    \node [sty,fill=convcolor] (conv2) [below=0.8mm of relu1] {Convolution\\[0mm]\tiny $(3\!\times\!)3\!\!\times\!\!3, C\!\rightarrow\!C$};

    \node [sty, fill=noisecolor, draw, densely dashed] (noise2) [below=0.8mm of conv2] {Noise Injection};
    \node [sty,fill=normcolor] (norm2) [below=0.8mm of noise2] {Normalization};
    \node [stysum, below=0.6mm of norm2] (sum) {$+$};
    \node [sty,fill=relucolor] (relu2) [below=0.6mm of sum] {Activation};

    \node [sty,fill=soundcolor] (sound) [minimum width=12mm, below left=0.1mm and 0.6mm of conv1.west] {Linear};
    \node [sty,fill=soundrelucolor] (soundrelu) [minimum width=12mm, below=0.6mm of sound.south] {Activation};
    
    \coordinate[above=2mm of sound.north] (soundIN);

    \draw[-, densely dashed]
        (soundIN) node [above, align=center] {Sound\\ Features} -- (sound);
    \draw[-]
        (sound) -- (soundrelu);
    \draw[->, densely dashed] (soundrelu) |- (norm1);
    \draw[->, densely dashed] (soundrelu) |- (norm2);

    \coordinate[above=2mm of sampling.north, label=right:Activations] (IN);
    \coordinate[below=3mm of relu2.south] (OUT);

    \draw[-]
        (IN) -- (sampling)
        (sampling) -- (conv1)
        (conv1) -- (noise1)
        (noise1) -- (norm1)
        (norm1) -- (relu1)
        (relu1) -- (conv2)
        (conv2) -- (noise2)
        (noise2) -- (norm2)
        (norm2) -- (sum)
        (sum) -- (relu2);
    \draw[->](relu2) -- (OUT);

    
    \node [sty,fill=convcolor, minimum width=3mm] (res) [right=0.6mm of norm1] {Conv.\\[0mm]\tiny $(1\!\times\!)1\!\!\times\!\!1$\\\tiny $C^{in}\!\!\!\rightarrow\!C$};

    \coordinate[below=0.8mm of sampling.south] (resIN);
    
    \draw[-, densely dashed] (resIN) -| (res.north);
    \draw[->, densely dashed] (res.south) |- (sum);

\end{tikzpicture}

%% file: diagrams/conv_block.tex
\tikzstyle{sty}=[minimum width=2.0cm, minimum height=0.4cm, font=\footnotesize, inner sep=3.0pt]
\begin{tikzpicture}[node distance=0.55cm,auto,>=latex']

    \node [sty,fill=teal!70] (conv) {Convolution};
    \node [sty,fill=teal!60] (noise) [below of=conv] {Noise Injection};
    \node [sty,fill=teal!45] (batch) [below of=noise] {Batch Norm};
    \node [sty,fill=teal!30] (relu) [below of=batch] {LeakyReLU};

    \coordinate[above=0.2cm of conv] (in);
    \coordinate[below=0.35cm of relu] (out);

        
    \path[-] (in) edge (conv);
    \path[-] (conv) edge (noise);
    \path[-] (noise) edge (batch);
    \path[-] (batch) edge (relu);
    \path[->] (relu) edge (out);

\end{tikzpicture}

%% file: diagrams/conv_layer.tex
\tikzstyle{sty}=[minimum width=2.5cm, minimum height=0.6cm, font=\footnotesize, inner sep=2.0pt]
\newcommand\subtitlefont{\footnotesize\linespread{0.6}\selectfont}
\begin{tikzpicture}[node distance=0.7cm,auto,>=latex']
    \node [sty,fill=red!30] (sampling) {$2\!\uparrow\!/2\!\downarrow$ Sampling};
    
    \node [sty,fill=purple!40] (conv1) [below=0.15cm of sampling, align=center, font=\subtitlefont] {Conv Block\\\tiny $(3\!\times\!)3\!\!\times\!\!3, C^{l-1}\!\rightarrow\!C^l$};
    
    \node [sty,fill=violet!40] (conv2) [below=0.15cm of conv1, align=center, font=\subtitlefont] {Conv Block\\\tiny $(3\!\times\!)3\!\!\times\!\!3, C^l\!\rightarrow\!C^l$};
    

    \node [sty,fill=brown!30] (res1) [right=0.15cm of conv1, align=center, minimum width=2.2cm, font=\subtitlefont] {Convolution\\\tiny $(1\!\times\!)1\!\!\times\!\!1,\!C^{l-1}\!\!\rightarrow\!\!C^l$};
        
    \node [sty,fill=brown!10] (res2) [right=0.15cm of conv2, align=center, minimum width=2.2cm, font=\subtitlefont] {LeakyReLU};
    
    
    \coordinate[below=0.07cm of sampling] (resIN);
    \coordinate[below=0.10cm of conv2] (resOUT);
    \coordinate[above=0.2cm of sampling] (in);
    \coordinate[below=0.35cm of conv2] (out);

    \path[-] (in) edge node {} (sampling);
    \path[-] (sampling) edge node {} (conv1);
    \path[-] (conv1) edge node {} (conv2);
    \path[->] (conv2) edge node {} (out);

    \draw[] (resIN) -| (res1.north) {};
    \path[] (res1) edge (res2);
    \draw[->] (res2.south) |- (resOUT) {};
    
\end{tikzpicture}

%% file: diagrams/gen_layers.tex
\tikzstyle{styconv}=[inner sep=2.5pt,fill=\gencolor]
\tikzstyle{styto}=[inner sep=2.5pt, fill=\gencolor]
\tikzstyle{styup}=[inner sep=2.0pt, fill=gray!30]

\begin{tikzpicture}[node distance=0.4cm, font=\scriptsize, auto,>=latex']
    
    \node (sum1) [stysum] {$+$};
    \node (up1)  [styup, below of=sum1] {Up};
    \node (sum2) [stysum, below of=up1] {$+$};
    
    \node (in) [above of=sum1] {};
    \node (out) [below=0.3cm of sum2] {};
    
    \node [styto] (to1) [right=0.3cm of sum1.east] {tRGB};
    \node [styto] (to2) [right=0.3cm of sum2.east] {tRGB};
    
    \node [styconv] (conv1) [right=0.2cm of to1.east] {Conv Layer};
    \node [styconv] (conv2) [right=0.2cm of to2.east] {Conv Layer};
    
    \path[-] (conv1) edge node {} (to1);
    \path[-] (conv2) edge node {} (to2);
    
    \path[->] (to1) edge node {} (sum1);
    \path[->] (to2) edge node {} (sum2);
    
    \path[-] (conv1) edge node {} (conv2);
    
    \path[-] (sum1) edge node {} (up1);
    \path[-] (up1) edge node {} (sum2);
    
    \path[-] (in) edge node {} (sum1);
    \path[->] (sum2) edge node {} (out);
    
    \node (convin) [above=0.3cm of conv1] {};
    \path[-] (convin) edge node {} (conv1);

    \node (convout) [below=0.3cm of conv2] {};
    \path[->] (conv2) edge node {} (convout);
    
    \node[above left=-0.05cm and 0.0cm of conv1.north, font=\tiny] (res) {$256\times 256$};
    \node[above left=-0.05cm and 0.0cm of conv2.north, font=\tiny] {$512\times 512$};

\end{tikzpicture}

%% file: diagrams/disc_layers.tex
\tikzstyle{styconv}=[inner sep=2.5pt,fill=\discolor]
\tikzstyle{styto}=[inner sep=2.5pt, fill=\discolor]
\tikzstyle{stydown}=[inner sep=2.0pt, fill=gray!30]
\tikzstyle{stysum}=[circle, draw=black, fill=white, minimum size=0.2cm, inner sep=-1pt, font=\tiny]

\begin{tikzpicture}[node distance=0.4cm, font=\scriptsize, auto,>=latex'] 

    \coordinate (in2) {};
    
    \coordinate [below of=in2] (c1);
    \node (down)  [stydown, below of=c1] {Down};
    \coordinate [below of=down] (c2);
    
    \node (out2) [below=0.3cm of c2, coordinate] {};
    
    \node [styto] (from1) [right=0.4cm of c1.east] {fRGB};
    \node [styto] (from2) [right=0.4cm of c2.east] {fRGB};
    
    \node [styconv] (conv3) [right=0.2cm of from1.east] {Conv. Layer};
    \node [styconv] (conv4) [right=0.2cm of from2.east] {Conv. Layer};
    
    \path[->] (from1) edge node {} (conv3);
    \path[->] (from2) edge node {} (conv4);
    
    \path[-] (c1) edge node {} (from1);
    \path[-] (c2) edge node {} (from2);
    
    \path[-] (conv3) edge node {} (conv4);
    
    \path[-] (in2) edge node {} (c1);
    \path[-] (c1) edge node {} (down);
    \path[-] (down) edge node {} (c2);
    \path[->] (c2) edge node {} (out2);
    
    \node (convin2) [above=0.3cm of conv3] {};
    \path[-] (convin2) edge node {} (conv3);

    \node (convout2) [below=0.3cm of conv4] {};
    \path[->] (conv4) edge node {} (convout2);
    
    \node[above left=-0.05cm and 0.0cm of conv3.north, font=\tiny] {$512\times 512$};
    \node[above left=-0.05cm and 0.0cm of conv4.north, font=\tiny] {$256\times 256$};
    
\end{tikzpicture}

%% file: diagrams/video_prediction.tex
\tikzstyle{sty}=[align=center, font=\scriptsize, inner sep=3.0pt]
\tikzstyle{stygru}=[anchor=north, align=center, font=\scriptsize, inner sep=2.5pt, minimum width=16mm, text depth=1.7mm]

\tikzstyle{c1}=[fill={rgb,255:red,245; green, 86; blue, 137}]
\tikzstyle{c2}=[fill={rgb,255:red,252; green, 214; blue, 125}]
\tikzstyle{c3}=[fill={rgb,255:red,35; green, 210; blue, 202}]
\tikzstyle{c4}=[fill={rgb,255:red,23; green, 193; blue, 237}]

\newcommand\ksize{0.17}
\tikzstyle{stykernel}=[minimum width=\ksize cm, minimum height=\ksize cm, inner sep=0pt]
\tikzstyle{k1}=[draw=blue!20,fill={rgb,255:red,7; green,0; blue,77}]
\tikzstyle{k2}=[draw=blue!20,fill=blue!20]

\newcommand{\makekernel}[9]{
    \node [stykernel, #1, below right=-(\ksize*1)+0.05 and (\ksize*1) ] {};
    \node [stykernel, #2, below right=-(\ksize*1)+0.05 and (\ksize*2) ] {};
    \node [stykernel, #3, below right=-(\ksize*1)+0.05 and (\ksize*3) ] {};
    \node [stykernel, #4, below right=(\ksize*0)+0.05 and (\ksize*1) ] {};
    \node [stykernel, #5, below right=(\ksize*0)+0.05 and (\ksize*2) ] {};
    \node [stykernel, #6, below right=(\ksize*0)+0.05 and (\ksize*3) ] {};
    \node [stykernel, #7, below right=(\ksize*1)+0.05 and (\ksize*1) ] {};
    \node [stykernel, #8, below right=(\ksize*1)+0.05 and (\ksize*2) ] {};
    \node [stykernel, #9, below right=(\ksize*1)+0.05 and (\ksize*3) ] {};
}

\tikzstyle{styhad}=[font=\large, circle, draw=black, fill=white, minimum size=2.0mm, inner sep=-1pt]

\begin{adjustbox}{width=\linewidth}
\begin{tikzpicture}[font=\footnotesize, auto, >=latex']

    \coordinate (IN) at (0,0);
    \coordinate[right=5mm of IN] (split1);
    
    \node [stygru,c2] (gru3) [right=2mm of split1] {ConvGRU\\\vspace{-1mm}\\$\qquad\quad V^c$};
    \node [stygru,c2] (gru2) [below=1.1mm of gru3] {DirConvGRU\\\vspace{-1mm}\\$\qquad\quad V^{v+}$};
    \node [stygru,c2] (gru1) [below=1.8mm of gru2] {DirConvGRU\\\vspace{-1mm}\\$\qquad\quad V^{v-}$};
    \node [stygru,c2] (gru4) [above=1.1mm of gru3] {DirConvGRU\\\vspace{-1mm}\\$\qquad\quad V^{h-}$};
    \node [stygru,c2] (gru5) [above=1.8mm of gru4] {DirConvGRU\\\vspace{-1mm}\\$\qquad\quad V^{h+}$};
    \node [sty] (x1) [below=0.5mm of gru1] {};

    \begin{scope}[shift={($(gru1.west)$)}]
        \makekernel{k2}{k2}{k2}{k2}{k1}{k2}{k1}{k1}{k1}
    \end{scope}
    \begin{scope}[shift={($(gru2.west)$)}]
        \makekernel{k1}{k1}{k1}{k2}{k1}{k2}{k2}{k2}{k2}
    \end{scope}
    \begin{scope}[shift={($(gru3.west)$)}]
        \makekernel{k1}{k1}{k1}{k1}{k1}{k1}{k1}{k1}{k1}
    \end{scope}
    \begin{scope}[shift={($(gru4.west)$)}]
        \makekernel{k2}{k2}{k1}{k2}{k1}{k1}{k2}{k2}{k1}
    \end{scope}
    \begin{scope}[shift={($(gru5.west)$)}]
        \makekernel{k1}{k2}{k2}{k1}{k1}{k2}{k1}{k2}{k2}
    \end{scope}
        
    \draw[to-to] (gru1) -- (gru2);
    \draw[to-to] (gru4) -- (gru5);

    \coordinate[right=5mm of gru1] (c1out);
    \coordinate[right=5mm of gru2] (c2out);
    \coordinate[right=5mm of gru3] (c3out);
    \coordinate[right=5mm of gru4] (c4out);
    \coordinate[right=5mm of gru5] (c5out);

    \node [stycat, right=8mm of gru3] (concat) {$\mdoubleplus$};
                
    \draw[-]
        (IN) -- (split1) node[midway] {$x_t$}
        (split1) |- (gru1.west)
        (split1) |- (gru2.west)
        (split1) -- (gru3.west)
        (split1) |- (gru4.west)
        (split1) |- (gru5.west)
        (split1) |- (x1.west);

    \path[-]
        (gru1.east) edge (c1out) node[below right] {$x_t^{v-}$}
        (gru2.east) edge (c2out) node[below right] {$x_t^{v+}$}
        (gru3.east) edge (c3out) node[above right] {$x_t^c$}
        (gru4.east) edge (c4out) node[above right] {$x_t^{h-}$}
        (gru5.east) edge (c5out) node[above right] {$x_t^{h+}$};
    \path[->]
        (c1out) edge (concat.280)
        (c2out) edge (concat.205)
        (c3out) edge (concat.180)
        (c4out) edge (concat.155)
        (c5out) edge (concat.90);      

    \coordinate[right=5mm of concat] (split2);
    
    \node [sty,c3] (wb) [above right=12mm and 1.5mm of split2] {Weighted Blending\\\tiny Conv $1\!\!\times\!\!1, 5P\!\!\rightarrow\!\!C$};
    
    \node [sty,c4] (mask) [below right=12mm and 1.5mm of split2, minimum width=21mm] {Attention Mask\\\tiny Conv $1\!\!\times\!\!1, 5P\!\!\rightarrow\!\!1$};
    
    \draw[-] (concat) -- (split2) node[above, pos=0.4] {$X_t$};
    \draw[-] (split2) |- (wb.west);
    \draw[-] (split2) |- (mask.west);

    \node [stysum, right=30mm of concat] (sum) {$+$};
    
    \coordinate[above=3.9mm of mask.north] (maskout);
    \node [styhad] (masking1) [below=6mm of sum] {$\cdot$};
    \node [styhad] (masking2) [above=6mm of sum] {$\cdot$};
    \node [sty, fill=none, draw=black, inner sep=2pt] (inv) [above=6.5mm of maskout] {$1-a$};

    \path[-] (mask.north) edge (maskout) node[above left] {$a_t$};
    \draw[-] (x1.west) -| (masking1) node[left, pos=0.49, label=$x_t$] {};
    \draw[->] (maskout) -- (masking1);
    
    \draw[-] (wb) -| (masking2) node[above, pos=0.4] {$x'_t$};
    \draw[-] (maskout) -- (inv);
    \draw[->] (inv) |- (masking2);

    \draw[->] (masking1.north) -- (sum); 
    \draw[->] (masking2.south) -| (sum); 

    \coordinate [right=4mm of sum] (OUT);
    \draw[->] (sum) -- (OUT) node [above, pos=0.7] {$x_t''$};
    
\end{tikzpicture}
\end{adjustbox}

%% file: tables/ablation.tex
\newcommand{\specialcell}[3][c]{%
    \rotatebox[origin=c]{90}{
  \begin{tabular}[#1]{@{}c@{}}\\[-6ex]\scriptsize#2\\[-1ex]\scriptsize#3\\[-6ex]\end{tabular}}
  }
\resizebox{1.0\linewidth}{!}
{
\begin{tabular}{lllllllllllllllll}
\cmidrule{3-17}
&
&\multicolumn{3}{c}{\textit{Cello}}     &
&\multicolumn{3}{c}{\textit{Quintet}}   &
&\multicolumn{3}{c}{\textit{Drums}}     &
&\multicolumn{3}{c}{\textit{Talking Head}}    \\
\cmidrule{3-5} \cmidrule{7-9} \cmidrule{11-13} \cmidrule{14-17}
&
& SSIM ($\uparrow$) & LPIPS ($\downarrow$)  & FVD ($\downarrow$) &
& SSIM ($\uparrow$) & LPIPS ($\downarrow$)  & FVD ($\downarrow$) &
& SSIM ($\uparrow$) & LPIPS ($\downarrow$)  & FVD ($\downarrow$) &
& SSIM ($\uparrow$) & LPIPS ($\downarrow$)  & FVD ($\downarrow$)  \\
\midrule
& Baseline (MfS)  & $0.82\pms0.06$ &	$0.16\pms0.06$ & $2219\pms1095$ &&  $0.69\pms0.11$ & $0.24\pms	0.10$ & $6480\pms2543$ &&  $0.68\pms0.02$ & $0.16\pms0.02$ & $2387\pms976$ && $0.43\pms0.03$    & $0.62\pms0.02$ & $1082\pms349$ \\
\cmidrule{2-17}
\multirow{3}{*}{\specialcell{Baseline}{+}}
& R+D-RNN       & $0.87\pms0.03$ & $0.09\pms0.02$ & $1184\pms905$  && $0.89\pms0.02$ & $0.05\pms0.02$ & $1706\pms2537$ && $0.75\pms0.03$ & $0.12\pms0.02$ & $2230\pms1276$ && $0.81\pms0.03$ & $0.07\pms0.02$ & $520\pms555$ \\
& SaC           & $0.87\pms0.03$ & $0.09\pms0.02$ & $1354\pms980$  && $0.89\pms0.02$ & $0.05\pms0.02$ & $1775\pms2701$ && $0.75\pms0.03$ & $0.12\pms0.02$ & $2127\pms1977$ && $0.77\pms0.03$ & $0.07\pms0.01$ & $366\pms470$ \\
& acIN          & $0.87\pms0.03$ & $0.08\pms0.02$ & $1426\pms1195$ && $0.88\pms0.02$ & $0.05\pms0.01$ & $1903\pms1866$ && $0.76\pms0.03$ & $0.11\pms0.03$ & $2122\pms2857$ && $0.69\pms0.02$ & $0.11\pms0.01$ & $486\pms362$ \\
\cmidrule{2-17}
\multirow{3}{*}{\vspace{2ex}\specialcell{All Abv.}{+}}
& ConvGRU       & $0.87\pms0.03$ & $0.08\pms0.02$ & $1101\pms1011$ && $0.90\pms0.02$ & $0.05\pms0.01$ & $1502\pms2090$ &&  $0.75\pms0.03$ & $0.11\pms0.02$ & $1700\pms1603$ && $0.77\pms0.03$ & $0.06\pms0.01$ & $327\pms435$ \\
& DirConvGRU    & $0.87\pms0.03$ & $0.08\pms0.02$ & $1085\pms991$ && $0.88\pms0.02$ & $0.06\pms0.02$ & $1455\pms1999$ &&  $0.75\pms0.03$ & $0.11\pms0.02$ & $1614\pms1390$ && $0.80\pms0.03$ & $0.05\pms0.01$ & $280\pms414$ \\

\bottomrule
\end{tabular}
}

%% file: tables/base_vs_full_FSD50K.tex
\begin{tabular}{lcccc}
\toprule
FID ($\downarrow$)
&\textit{Cello}
&\textit{Quintet}
&\textit{Drums}
&\textit{Talking Head}
\\
\midrule
Baseline (MfS)  & $320\pms14$ & $423\pms20$ & $385\pms14$ & $528\pms16$ \\
Full Model      & $279\pms24$ & $405\pms29$ & $378\pms28$ & $535\pms42$ \\
\bottomrule
\end{tabular}

%% file: figures/synth_seq_results.tex
    \includegraphics[height=0.15\linewidth, width=0.07\linewidth]{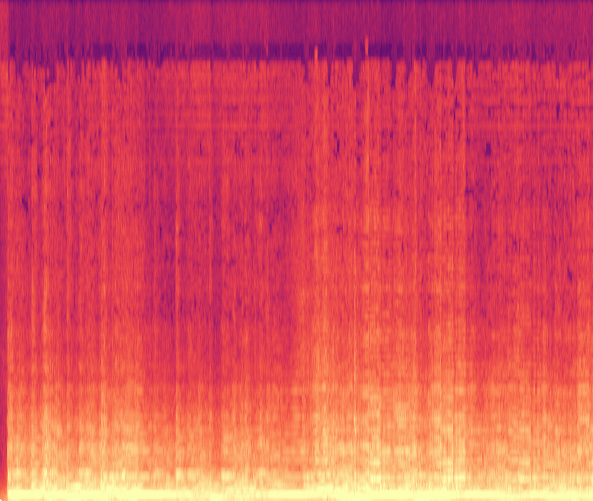}
    \includegraphics[width=0.15\linewidth]{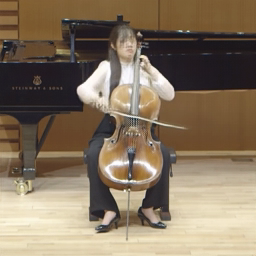}
    \includegraphics[width=0.15\linewidth]{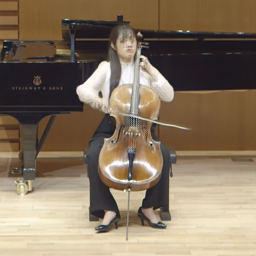}
    \includegraphics[width=0.15\linewidth]{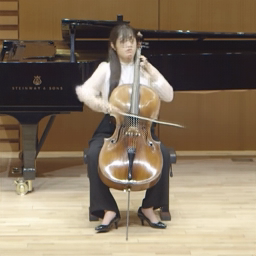}
    \includegraphics[width=0.15\linewidth]{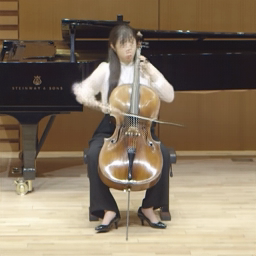}
    \includegraphics[width=0.15\linewidth]{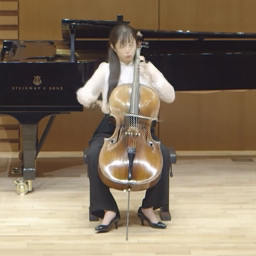}
    \includegraphics[width=0.15\linewidth]{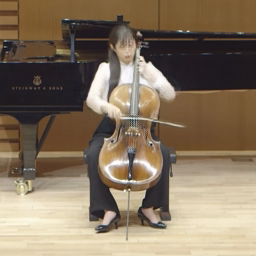}
    \\
    \includegraphics[height=0.15\linewidth, width=0.07\linewidth]{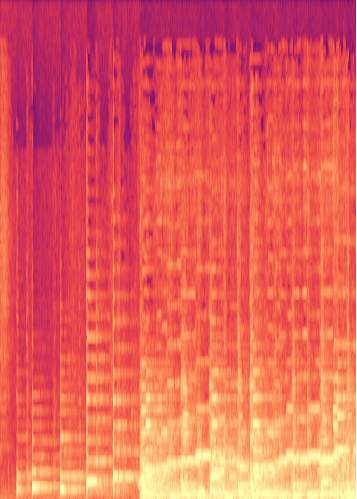}
    \includegraphics[width=0.15\linewidth]{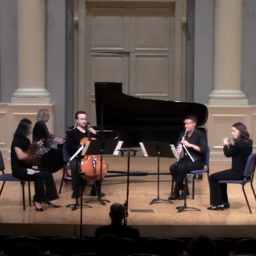}
    \includegraphics[width=0.15\linewidth]{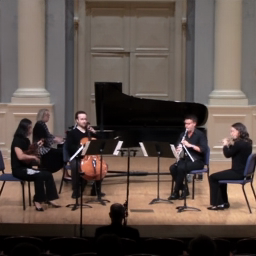}
    \includegraphics[width=0.15\linewidth]{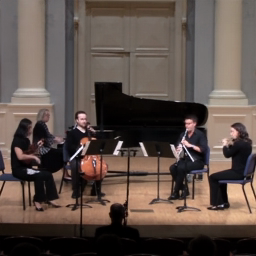}
    \includegraphics[width=0.15\linewidth]{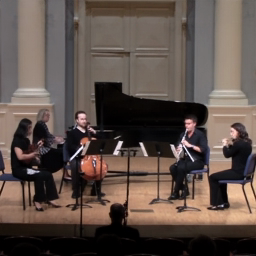}
    \includegraphics[width=0.15\linewidth]{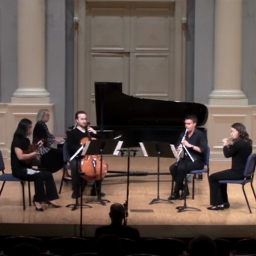}
    \includegraphics[width=0.15\linewidth]{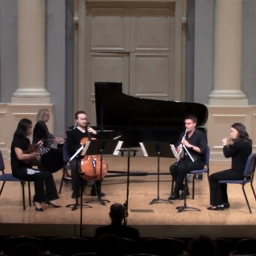}
    \\
    \includegraphics[height=0.15\linewidth, width=0.07\linewidth]{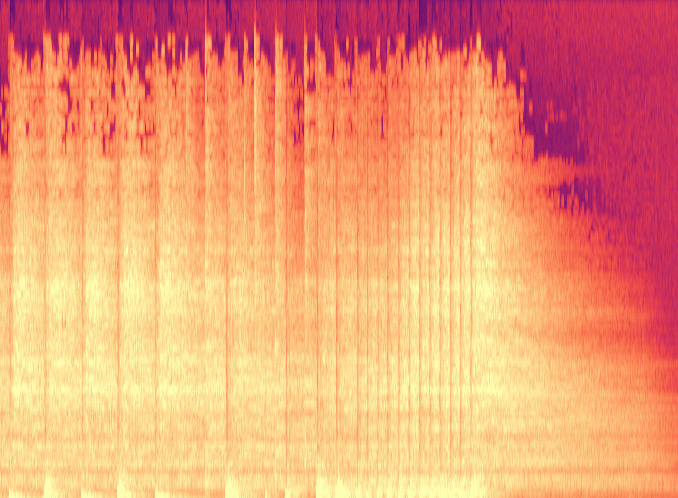}
    \includegraphics[width=0.15\linewidth]{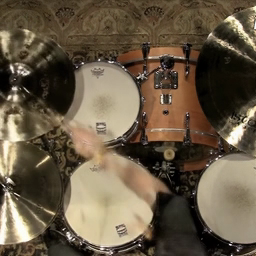}
    \includegraphics[width=0.15\linewidth]{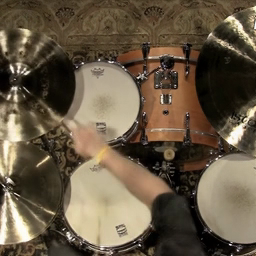}
    \includegraphics[width=0.15\linewidth]{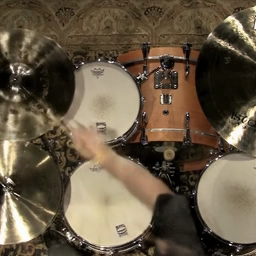}
    \includegraphics[width=0.15\linewidth]{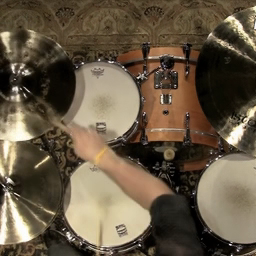}    
    \includegraphics[width=0.15\linewidth]{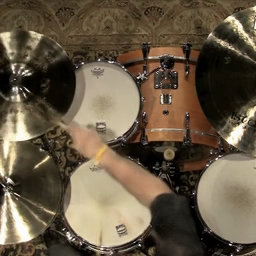}
    \includegraphics[width=0.15\linewidth]{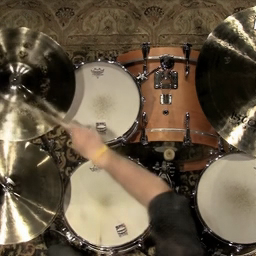}    
    \\
    \includegraphics[height=0.15\linewidth, width=0.07\linewidth]{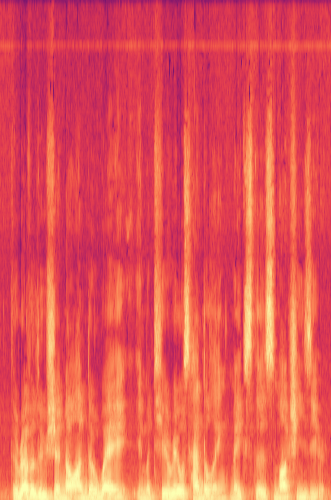}
    \includegraphics[width=0.15\linewidth]{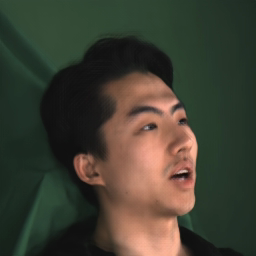}
    \includegraphics[width=0.15\linewidth]{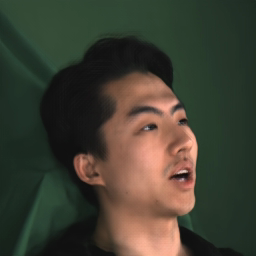}
    \includegraphics[width=0.15\linewidth]{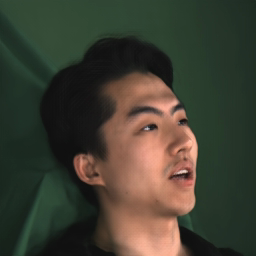}
    \includegraphics[width=0.15\linewidth]{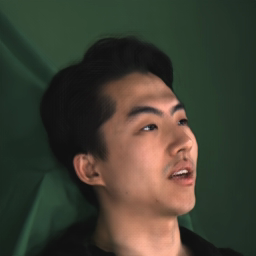}
    \includegraphics[width=0.15\linewidth]{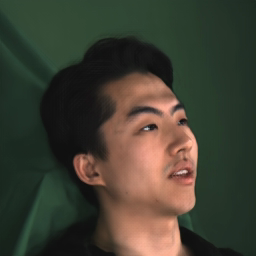}
    \includegraphics[width=0.15\linewidth]{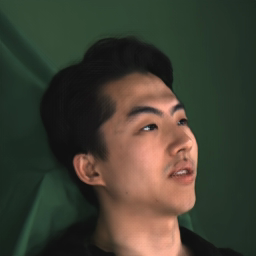}